\documentclass[a4paper,12pt]{article}
\usepackage{cite}
\usepackage{epsfig, amsmath, amssymb}
\usepackage{graphicx,psfrag}
\usepackage{xcolor} 
\usepackage{graphicx}
\usepackage{amssymb}
\usepackage{amsmath}
\usepackage{amsfonts}
\usepackage{dsfont}
\usepackage{mathtools}
\usepackage{slashed}
\usepackage{amsmath}
\usepackage{caption}
\usepackage{tikz}
\usepackage{rotating}
\usepackage{bbold,amsfonts}
\usepackage{caption}
\usepackage{subcaption}

\usepackage[utf8]{inputenc}
\usepackage{bm}
\usepackage{xcolor}
\usepackage{float}
\usepackage{braket}
\numberwithin{equation}{section}
\usepackage[height=9.1in,width=6.45in]{geometry}
\usepackage[font=small,labelfont=bf]{caption}
\usepackage[hidelinks]{hyperref}
\usepackage{mathtools}
\usepackage{multirow}

\newcommand\vertarrowbox[3][6ex]{%
	\begin{array}[t]{@{}c@{}} #2 \\
		\left\downarrow\vcenter{\hrule height #1}\right.\kern-\nulldelimiterspace\\
		\makebox[0pt]{\scriptsize#3}
	\end{array}%
}

\title{$\kappa$-symmetric M5 brane web for defects in $AdS_7$/$CFT_6$ holography}
\date{February 2025}

\begin{document}
\begin{titlepage}
		\vbox{
			\halign{#\hfil         \cr
			} 
		}  
		\vspace*{15mm}
		\begin{center}
			{\Large \bf 
				Linking defects via $AdS$/$CFT$ holography
			}
			
			\vspace*{15mm}
			
			{\large Varun Gupta}
			\vspace*{8mm}
			
			COEP Technological University, \\
			Shivaji Nagar, Pune, Maharashtra 411005, India \\
			
			\vskip 0.8cm
			
			
			{\small
				E-mail:  vjg.physics@coeptech.ac.in
			}
			\vspace*{0.8cm}
		\end{center}
		
		\begin{abstract}

            \noindent

          In this article, we discuss some supersymmetric probe D5 brane solutions embedded in a type 2b supergravity background solution obtained by deforming the global $AdS_5 \times S^5$ spacetime. The D5 brane solution of interest has a non-compact worldvolume, and when it is placed alongside overlapping with an  anti-$\overline{\text{D5}}$ solution, the combination is dual to a codimension-1 defect of the boundary $SYM$ theory. The dual defect becomes a topological symmetry defect of a $U(1)$ $0$-form global symmetry if we move the profile of the probe D5 more towards the boundary of the bulk spacetime, resulting in a U-shaped profile hanging from the boundary. We show how the hanging D5 brane can be used holographically to measure the $U(1)$ global symmetry charge of the heavy determinant operators in the SYM theory by moving them across dual giant graviton D3 branes in the bulk spacetime.

		\end{abstract}
		\vskip 1cm
		{
		}
	\end{titlepage}
	 

\section{Introduction}

In quantum field theories and gauge theories, defects now occupy a very special place. In the last decade, they have been very important in understanding the dynamics of the theory in several ways. They have been useful in probing the theory in ways that local operators cannot. Over the last decades, they have enhanced our understanding of symmetries, dualities, confinement, anomalies and conformal field theories \cite{Gukov:2014, Gukov:2006jk, Gukov:2008sn, GGS:2013, FGT:2015, Leflochetal:0217, SKAMBetal:0717, SKAMBetal:0718, GKSW:2014, Tseytlinetal:2017, Jensenetal:2018, GN:2022, Zaremboetal:2021, Holguin:2503, Chalabietal:2503, Billoetal:2016, Wangetal2020}. Defects probe objects like monopoles, vortices, instantons, etc. that are invisible in perturbation theory \cite{GGS:2013, FGT:2015, Leflochetal:0217, SKAMBetal:0717, SKAMBetal:0718}. Certain defects preserve the integrable structure of the bulk theory; examples include defect spin chains and integrable boundary conditions, etc. \cite{Zaremboetal:2021, Holguin:2503, Chalabietal:2503}. In defect conformal field theory, a codimension-q defect preserves an $SO(d-q+1,1) \times SO(q)$ subgroup of the $SO(d+1,1)$ conformal symmetry, which leads to rich new observables, including defect operator expansions, displacement operators, defect central charges, etc. \cite{Billoetal:2016, Tseytlinetal:2017, Jensenetal:2018, Wangetal2020}. In the last ten years or so, they have been useful in revealing generalised global symmetries starting from the work of Gaiotto, Kapustin, Seiberg, and Willett in reference \cite{GKSW:2014}, where generalised symmetries were formulated in terms of topological defects and have ever since reshaped our understanding of symmetries in QFT. See the references \cite{Bhardwajetal:2023, Sakura:2023, Kaidi:2026, Bahetal:2022, Etxebarria:2022, Heckmanetal:2022, Bergmanetal:2024, Beninietal:2024, Waddleton:2024, Mignosaetal:2025, Bahetal:2025, Bahetal:2026, Mignosaetal:2026, Rodriguez-Gomez:2026} and references therein for details on the developments.

According to $AdS/CFT$ correspondence, the gauge symmetries of the bulk supergravity theory are mapped to the global symmetries of the boundary field theory \cite{Witten:1998, Maldacena:98}. And the global symmetries in QFTs are associated with some codimension-$q$ topological defects. In this article, we study the holography of codimension-1 defects in the 4d $\mathcal{N}=4$ SYM theory by looking at some probe D5 branes embedded in a type 2b background solution which is a deformation of $AdS_5 \times S_5$ by turning on a U(1) gauge field connection \cite{Gauntlettetal:2003, Gauntlettetal:2004}. 
We will be interested in the case when the defect is topological and associated with some global symmetry of the boundary field theory.
However, the holographic duality of the topological defects is not so straightforward. 
Establishing the duality between the topological defects associated with discrete global symmetries have required symmetry topological field theory({\it sym-TFT}) methods by focusing on the topological sector of the supergravity theory and the dual probe D branes with the dynamics in their DBI action term frozen \cite{Bahetal:2022, Etxebarria:2022, Heckmanetal:2022}.
For continuous symmetries, the authors in \cite{Bergmanetal:2024}, based on the {\it sym-TFT}-based study done in the paper \cite{Beninietal:2024}, made the proposal that a codimension-$q$ top. defect associated with a $(d-q-1)$-form continuous global symmetry will be $AdS/CFT$ dual to some stable non-BPS D$p$ brane which lies very, very close to the boundary of $AdS$ and perpendicular to the radial direction. The linking number between the non-BPS D$p$ brane and any BPS D-brane(dual to the $d-q-1$ dimensional operator charged under this $(d-q-1)$-form symmetry) would measure the charge of the $d-q-1$ dimensional operator under this global symmetry. The authors in the paper \cite{Mignosaetal:2025}, based on the work in \cite{Sen:1998, Alishahiha:2000}, further suggested that the non-BPS D$p$-brane associated with the $(d-q-1)$-form global symmetry could be resolved into a D$(p+1)$-$\overline{\text{D}(p+1)}$-brane pair, whose profile look like a 'U' shape hanging from the $AdS$ boundary. In the last one year, the authors in the references \cite{Bahetal:2025, Bahetal:2026} made this proposal concrete by studying some U-shaped hanging D5 branes in the $AdS_5 \times S^5$ spacetime, which are associated with the 0-form global symmetry which is the R-symmetry of the $\mathcal{N}=4$ SYM of the non-Abelian group $SU(4)$.

In this article, our goal is to show that how if we consider our probe D5 brane solution, which are {\it holomorphically} embedded in the global $AdS_5 \times S^5$ spacetime, and tune the embedding parameter in a certain way, then we can make their profile look like hanging U-shaped D5 branes of \cite{Mignosaetal:2025, Bahetal:2025, Bahetal:2026}. The hanging probe D5 when placed alongside its anti $\overline{\text{D5}}$ partner -- overlaping in 4 spacelike directions -- should be equivalent to the non-BPS D4 brane which is dual to codimension-1 top. defect of U(1) global symmetry. In section 2, we start off by presenting the type 2b background solution we work with. We derive a general BPS embedding condition of probe D5 branes by solving the $\kappa$-symmetry equation. Since this result is very general, we consider a simpler D5 embedding solution which has noncompact worldvolume and stretches along the radial direction to end on the boundary. We analyze this solution by tuning the real constant parameter in the embedding condition. We focus in the limit when the profile of the probe D5 hangs out slighly away from the boundary in a 'U' shape. We calculate the effective action of this probe solution. Further, we also present a general solution for anti-$\overline{\text{D5}}$ brane by solving the $\kappa$-symmetry equation with a minus sign. In section 3, we present a general BPS embedding condition for probe D3 branes in this background. This general solution has already been derived previously in \cite{AS:2008} in $AdS \times S^5$ and looked into in the context of codimension-2 defects in \cite{AGS:2020}. We will be focusing on the subset of the embedding condition of giant graviton D3 branes of \cite{Mikhailov:2000, Grisaru:2000zn, McGreevy:2000cw}. The holographic duals of the giants would be the heavy determinant operators in the boundary theory charged under the $U(1)$ 0-form global symmetry. In subsection \ref{chargemeasurement}, following \cite{Bahetal:2022, Bahetal:2025, Bahetal:2026}, we will discuss how we can explain the charge measurement of the determinant operator via Hanany-Witten transition by shifting the giant graviton D3 brane and the D5-$\overline{\text{D5}}$ brane pair relative to each other \cite{HananyWitten:96}. In section 4 we will conclude with the discussion of results and future directions.

\section{New Probe D5 solutions}\label{section2}

In this section we work in the supergravity background solution which was derived in the paper \cite{Gauntlettetal:2004}. It can be obtained from the global $AdS_5 \times S^5$ solution by turning on a deformation parameter $f$ which also turns on the $U(1)$ gauge field $A$ in this background. More interestingly this solution leads to closed time like trajectory curve if the deformation parameter $f$ becomes greater than a critical value. This solution was obtained as a part of classification program of supersymmetric solutions of five-dimensional gauged supergravity which started in the work \cite{Gauntlettetal:2003}.

\subsection{A Type 2b solution deformed from $AdS_5 \times S^5$}\label{deformedAdS}

The metric solution is given by
\begin{equation}
    ds_{10}^2 \, = \, ds_5^2 \, + \, l^2 \sum_{i=1}^3 \left[ \left(d\mu_i \right)^2 \, + \, \mu_i^2 \left( d\xi_i \, + \, \frac{2}{l \sqrt{3}} A \right)^2 \right]
\end{equation}
where $\mu_1 = \sin \alpha$, $\mu_2 = \cos \alpha \sin \beta$, $\mu_3 = \cos \alpha \cos \beta$ with $0 \le \alpha \le \pi/2$, $0 \le \beta \le \pi/2$, $0 \le \xi_i \le 2\pi$ and together they parametrises an $S^5$. \\
\noindent
Here
\begin{equation}
    ds_{5}^2 \, = \, - \left(dt + \frac{r^2}{2l} \sigma_3^L \, + \, \frac{f r^2}{V(r)} \sigma_1^L  \right)^2 \, + \, \frac{dr^2}{V(r)} + \frac{r^2}4 \left[ \left( \sigma_1^L \right)^2 + \left( \sigma_2^L \right)^2 + V(r) \left( \sigma_3^L \right)^2 \right]
\end{equation}
with $V(r) = 1 + \frac{r^2}{l^2}$ and $\sigma_i^L$ are the right -invariant 1-forms on the three-sphere
\begin{align}\label{rightinvariantforms}
    \sigma_1^L &= \sin \phi d\theta - \sin \theta \cos \phi d\psi \cr
    \sigma_2^L &= \cos \phi d\theta + \sin \theta \sin \phi d\psi \cr
    \sigma_3^L &=  d\phi + \cos \theta  d\psi
\end{align}
$ds_{5}^2$ reduces to an $AdS_5$ metric when $f=0$ and $\phi$ is reparametrized to
\begin{equation}
    \tilde{\phi} = \phi - 2t/l\,,
\end{equation}
and $ds_{5}^2$ simplifies to
\begin{equation}
ds_{5}^2 \bigg|_{f \rightarrow 0 } \, = \,  - V(r) dt^2 + \frac{dr^2}{V(r)} + \frac{r^2}4 \left[ \left( \tilde{\sigma}_1^L \right)^2 + \left( \tilde{\sigma}_2^L \right)^2 +  \left( \tilde{\sigma}_3^L \right)^2 \right]
\end{equation}
where $\tilde{\sigma}_i^L$ are as in \eqref{rightinvariantforms} with $\phi$ replaced with $\tilde{\phi}$.\\

\noindent
This type 2b solution also includes a non-zero 5-form flux field given by
\begin{equation}\label{selfdual5form}
    F^{(5)} \, = \, \left(1 + \star_{10} \right) \left[ - \frac 4l \text{vol}_{(5)} + \frac{l^2}{\sqrt{3}} \sum_{i=1}^3 d\left( \mu_i\right)^2 \wedge d\xi_i \wedge \star_{(5)} F^{(2)}\right]
\end{equation}
where the 2-form $F^{(2)}$ is the field strength of the $U(1)$ gauge field $A = \left( \sqrt{3}/2 \right) \left( \frac{f r^2}{V(r)} \sigma_1^L \right)$.\\

\noindent
The above solution preserves two super symmetries as shown in \cite{Gauntlettetal:2004}. We consider the following frame vielbein from this reference
\begin{align}
e^0 \, = \, dt + \frac{r^2}{2l} \sigma_3^L + \frac{f r^2}{V(r)} \sigma_1^L \,, \qquad \qquad e^1 \, = \,  \frac1{V^{\frac12}} dr \cr
e^2 \, =\, \frac r2 \sigma_1^L \,, \qquad \qquad e^3 \, =\, \frac r2 \sigma_2^L \,, \qquad \qquad e^4 \, =\, \frac r2 V^{\frac 12} \sigma_3^L \,,
\end{align}
\begin{align}
e^5 \, &= \, l d\alpha  \,, \qquad \qquad e^6 \, = \,  l \cos \alpha d\beta \,, \cr
e^7 \, &= \, l \sin \alpha \cos \alpha \left[ d\xi_1 - \sin^2 \beta d\xi_2 - \cos^2 \beta d\xi_3 \right]\,, \cr
e^8 \, &=  \, l \cos \alpha \sin \beta \cos \beta \left[d\xi_2 - d\xi_3 \right]\,, \cr
e^9 \, &= \, - \frac2{\sqrt{3}} A - l \sin^2 \alpha \, d\xi_1 - \cos^2 \alpha \left( \sin^2 \beta \, d\xi_2 + \cos^2 \beta \, d\xi_3 \right)
\end{align}
The constraint on the Killing spinor comes from the integrability conditions, where the following four projections need to be imposed on the Killing spinor
\begin{align}\label{1by16projections}
    \Gamma_{14} \, \epsilon \, = \, - i \, \epsilon & \qquad  \qquad \Gamma_{23} \, \epsilon \, = \,  i \, \epsilon \cr
    \Gamma_{57} \, \epsilon \, = \, - i \, \epsilon & \qquad  \qquad \Gamma_{68} \, \epsilon \, = \, - i \, \epsilon
\end{align}
where $\epsilon = \epsilon_1 + i \epsilon_2$ and  $\epsilon_1,\, \epsilon_2$ are both Majorana-Weyl spinors with
\begin{equation}
    \Gamma_{11} \, \epsilon \,=\, - \epsilon\,. 
\end{equation}
The chiral Killing spinor solution satisfying the projections \eqref{1by16projections} is given by
\begin{equation}
    \epsilon \, = \, e^{\frac i2 \left( \frac{3t}l - \xi_1 - \xi_2 - \xi_3 \right)} \, \eta
\end{equation}
where $\eta$ is a constant spinor satisfying all the projections.
\subsection{Finding the probe D5 solution in this background}\label{generalprobeD5}

We start with solving the $\kappa$-symmetry equation for the probe D5 given by
\begin{equation}\label{kappasymmetryeqn}
    \Gamma_{\kappa} \, \epsilon \, = \, \sqrt{- \text{det} \, h} \, \epsilon
\end{equation}
where $h$ is the induced metric on the D5. \\

\noindent
We start with the most general possible ansatz for the probe D5 and consider the following pull-backs of the 10d spacetime vielbein
\begin{equation}
    \mathfrak{e}^a_{i} \, = \, e^a_{\mu} \partial_i X^{\mu}\,,
\end{equation}
here index $i$ denotes the worldvolume coordinates $i \in \{\tau, \sigma_1, \ldots, \sigma_5 \}$, index $\mu$ denotes 10d spacetime coordinates. $\Gamma_{\kappa}$ in the equation \eqref{kappasymmetryeqn} is the pullback 
\begin{equation}
    \Gamma_{\kappa} \, = \, \mathfrak{e}^a \wedge \mathfrak{e}^b \wedge \mathfrak{e}^c \wedge \mathfrak{e}^d \wedge \mathfrak{e}^e \wedge \mathfrak{e}^f \, \Gamma_{abcdef}
\end{equation}
with $\Gamma_a$s being the 10d $\Gamma$ matrices. \\

\noindent
We follow the method first employed in the reference \cite{AGS:2020} and obtain the following 6-form BPS constraints
\begin{align}\label{6formconstraints}
    \mathfrak{e}^{09} \wedge \left( \omega \, + \, \widetilde{\omega} \right) \wedge \mathbf{E}^a \wedge \mathbf{E}^b \, = \, 0 & \qquad \qquad  \mathfrak{e}^{09}  \wedge \mathbf{E}^1 \wedge \mathbf{E}^2 \wedge \mathbf{E}^5 \wedge \mathbf{E}^6\, = \, 0 \cr
    \left( \mathfrak{e}^0 \, + \, \mathfrak{e}^9 \right) \wedge \mathbf{E}^a \wedge \left( \omega \, + \, \widetilde{\omega} \right) \wedge \left( \omega \, + \, \widetilde{\omega} \right) \, = \, 0 & \qquad \qquad \left( \mathfrak{e}^0 \, + \, \mathfrak{e}^9 \right) \wedge \mathbf{E}^a \wedge \mathbf{E}^b \wedge \mathbf{E}^c  \wedge \left( \omega \, + \, \widetilde{\omega}  \right) \, = \, 0 \cr
    \mathbf{E}^a \wedge \mathbf{E}^b  \wedge \left( \omega \, + \, \widetilde{\omega}  \right) \wedge \left( \omega \, + \, \widetilde{\omega}  \right) \, = \,0 & \qquad \qquad \left( \omega \, + \, \widetilde{\omega}  \right) \wedge \left( \omega \, + \, \widetilde{\omega}  \right) \wedge \left( \omega \, + \, \widetilde{\omega}  \right) \, =\,0 \cr
\end{align}
Where $\mathbf{E}^a$ are the complexified 1-forms
\begin{equation}\label{complexifiedoneform}
    \mathbf{E}^1 = \mathfrak{e}^1 - i \,  \mathfrak{e}^4 \quad \qquad  \mathbf{E}^2 = \mathfrak{e}^2 + i \,  \mathfrak{e}^3 \quad \qquad \mathbf{E}^5 = \mathfrak{e}^5 - i \,  \mathfrak{e}^7 \quad \qquad \mathbf{E}^6 = \mathfrak{e}^6 - i \,  \mathfrak{e}^8
\end{equation}
and $\omega$ and $\widetilde{\omega}$ are the 2-forms
\begin{align}\label{kaehlerforms}
    \omega \, = \, \mathfrak{e}^{57} \, + \, \mathfrak{e}^{68} \qquad \qquad  \widetilde{\omega} \, = \, \mathfrak{e}^{14} \, - \, \mathfrak{e}^{23}\,.
\end{align}
Here $\omega$ would be a K\"{a}hler form on the base manifold $\mathbb{CP}^2$ in an $S^5$ when $f$ deformation is 0. They are related to $\mathbf{E}^a$ and their complex conjugate $\overline{\mathbf{E}}^a$ by the equation
\begin{align}
    \omega \, = \, \frac{- i}2 \left( \mathbf{E}^5 \overline{\mathbf{E}}^5 \, + \, \mathbf{E}^6 \overline{\mathbf{E}}^6 \right) \qquad \qquad \widetilde{\omega} \, = \, \frac{- i}2 \left( \mathbf{E}^1 \overline{\mathbf{E}}^1 \, + \, \mathbf{E}^4 \overline{\mathbf{E}}^4 \right)
\end{align}
\noindent
These are in total $6 + 1 + 4 + 4 + 6 + 1 = 22$ number of 6-form constraints in \eqref{6formconstraints}. \\

\noindent
The $\kappa$-symmetry equation also gives the 6-form volume form on the D5 worldvolume
\begin{equation}
    \text{dvol}_6 \, = \, \mathfrak{e}^{09} \wedge \left( \omega \, + \, \widetilde{\omega}  \right) \wedge \left( \omega \, + \, \widetilde{\omega}  \right)
\end{equation}
Since the six-dimensional worldvolume solution embedded in the 10d will be given by 4 real functional conditions, we consider the following two arbitrary complexified conditions for our general solution
\begin{align}\label{generalconditions}
    F^{(I)}\left(\phi_0, \phi_1, \phi_2, \rho, \theta, \alpha, \beta, \xi_1, \xi_2, \xi_3 \right) \, = \, 0 \qquad \qquad \qquad \left(\text{with} \,\, I= 1,2\right)
\end{align}
Here, we have considered the reparametrisation of the 10d coordinates in the following way
\begin{equation}
   t \, = \, l \, \phi_0 \qquad \tilde{\phi} \, =\, - \left( \phi_1 + \phi_2 \right) \qquad \psi \, =\, - \left( \phi_1 - \phi_2 \right) \quad \text{and} \quad \theta \rightarrow 2 \theta
\end{equation}
so that in the undeformed version, when $f=0$, the 10d $AdS_5 \times S^5$ spacetime can be parametrized using the following complexified coordinates
\begin{align}
\Phi_0 \, = \, l \cosh \rho \, e^{i \phi_0} & \qquad \Phi_1 \, = \, l \sinh \rho \cos \theta \,  e^{i \phi_1}  \qquad \Phi_2 \, = \, l \sinh \rho \sin \theta \,  e^{i \phi_2} \cr
Z_1 \, = \, l \sin \alpha \, e^{- i \xi_1} & \qquad Z_2 \, = \, l \cos \alpha \sin \beta \, e^{- i \xi_2}  \qquad Z_3 \, = \, l \cos \alpha \cos \beta \, e^{- i \xi_3}
\end{align}
After this, we can write the following differential conditions from \eqref{generalconditions}
\begin{align}\label{generaldifferentialconditions}
    \text{P} \left[  F_{\rho} d\rho \, + \, F_{\theta} d\theta \, + \, F_{\alpha} d\alpha  \, + \, F_{\beta} d\beta \, + \, \sum_{i=0}^2 F_{\phi_i} d\phi_i \, + \, \sum_{i=1}^3 F_{\xi_i} d\xi_i\right] \, = \, 0
\end{align}
where P denotes pullback onto the world-volume, the notation $F_i$ means $\partial_i F$, and $r = l \sinh \rho$. We write each of the 1-form differentials in terms of the complexified 1-forms introduced in equation \eqref{complexifiedoneform}
\begin{align}\label{BPSdifferentialconstraints}
 &{\small \left[  F^I_{\rho} - i \coth \rho \left( F^I_{\phi_1} + F^I_{\phi_2}\right)   - i  \tanh \rho \,F^I_{\phi_0}   \right] \mathbf{E}^1 } 
 +  \left[ F^I_{\rho}  + i \coth \rho \left( F^I_{\phi_1} + F^I_{\phi_2}\right)   + i \tanh \rho  F^I_{\phi_0}   \right] \overline{\mathbf{E}}^1 \cr
 &\small +  e^{\rho}\left(\coth \rho - 1 \right) \Big[ \left( \tan \theta F^I_{\phi_1} - \cot \theta  F^I_{\phi_2}  - i  F^I_{\theta} \right)  e^{2i  \phi_0 - i \left(\phi_1 + \phi_2\right)} - 2 f  l  \left(\sum_{i=0}^2 F^I_{\phi_i} + \sum_{i=1}^3 F^I_{\xi_i} \right)  \Big] \mathbf{E}^2 \cr
 &\small +  e^{\rho}\left(\coth \rho - 1 \right) \Big[ \left( \tan \theta  F^I_{\phi_1} - \cot \theta  F^I_{\phi_2}  + i  F^I_{\theta} \right)  e^{i \left(\phi_1 + \phi_2\right) -2i  \phi_0 } - 2 f  l  \left(\sum_{i=0}^2 F^I_{\phi_i} + \sum_{i=1}^3 F^I_{\xi_i} \right)  \Big] \overline{\mathbf{E}}^2 \cr
&\small +  \left[ F^I_{\alpha}  + i \cot \alpha   F^I_{\xi_1}   -  i \tan \alpha  \left(F^I_{\xi_2} + F^I_{\xi_3} \right)  \right]\mathbf{E}^5   + \left[ F^I_{\beta}  + \, i \, \cot \beta \, F^I_{\xi_2}   - \, i \, \tan \beta \, F^I_{\xi_3}  \right] \mathbf{E}^6  \cr
 &\small +  \left[ F^I_{\alpha}  - i \cot \alpha   F^I_{\xi_1}   +  i \tan \alpha  \left(F^I_{\xi_2} + F^I_{\xi_3} \right)  \right] \overline{\mathbf{E}}^5   +   \left[ F^I_{\beta}  - \, i \, \cot \beta \, F^I_{\xi_2}   + \, i \, \tan \beta \, F^I_{\xi_3}  \right] \overline{\mathbf{E}}^6 \cr
  &\small + \,  \left[ \left( \sum_{i=0}^2 \partial_{\phi_i} - \sum_{i=1}^3 \partial_{\xi_i} \right) F^I \right] \left(\mathfrak{e}^0  +  \mathfrak{e}^9 \right)   +  \left[ \left( \sum_{i=0}^2 \partial_{\phi_i} + \sum_{i=1}^3 \partial_{\xi_i} \right) F^I \right] \left(\mathfrak{e}^0  -  \mathfrak{e}^9 \right) \, = \, 0
\end{align}
We can solve the 22 number of 6-form BPS constraints shown in equation \eqref{6formconstraints} using the above equation \eqref{BPSdifferentialconstraints}, if we set the coefficients of $\overline{\mathbf{E}}^a$ and $\left( \mathfrak{e}^0 - \mathfrak{e}^9\right)$ to 0.\footnote{The constraints in equation \eqref{6formconstraints} are solved by subsitituting one of the $\mathbf{E}^a$ from \eqref{BPSdifferentialconstraints} in terms of other $\mathbf{E}^b, \overline{\mathbf{E}}^b$ and $\left(\mathfrak{e}^0  +  \mathfrak{e}^9 \right)$, etc. And in order to make equation \eqref{BPSdifferentialconstraints} consistent with BPS constraints in \eqref{6formconstraints} we will have to set we set the coefficients of $\overline{\mathbf{E}}^a$ and $\left( \mathfrak{e}^0 - \mathfrak{e}^9\right)$ to 0.}   \\

\noindent
And therefore, by solving these four coefficients equations in \eqref{BPSdifferentialconstraints}, we obtain our general embedding conditions, which are given by
\begin{equation}\label{generalprobeD5solution}
    F^I \left(\Phi_0, \Phi_1, \Phi_2, Z_1, Z_2, Z_3 \right) \, = \, 0 \, \qquad \qquad \left(\text{with} \,\, I= 1,2\right),
\end{equation}
which are two arbitrary holomorphic functional conditions, also satisfying the scaling condition given below
\begin{equation}
    \sum_{i=0}^2 \partial_{\phi_i} F^I + \sum_{i=1}^3 \partial_{\xi_i} F^I \, = \, 0 \,.
\end{equation}
\begin{center}
	\begin{figure}[ht]
\begin{center}\includegraphics[width=37pc,height=18pc]{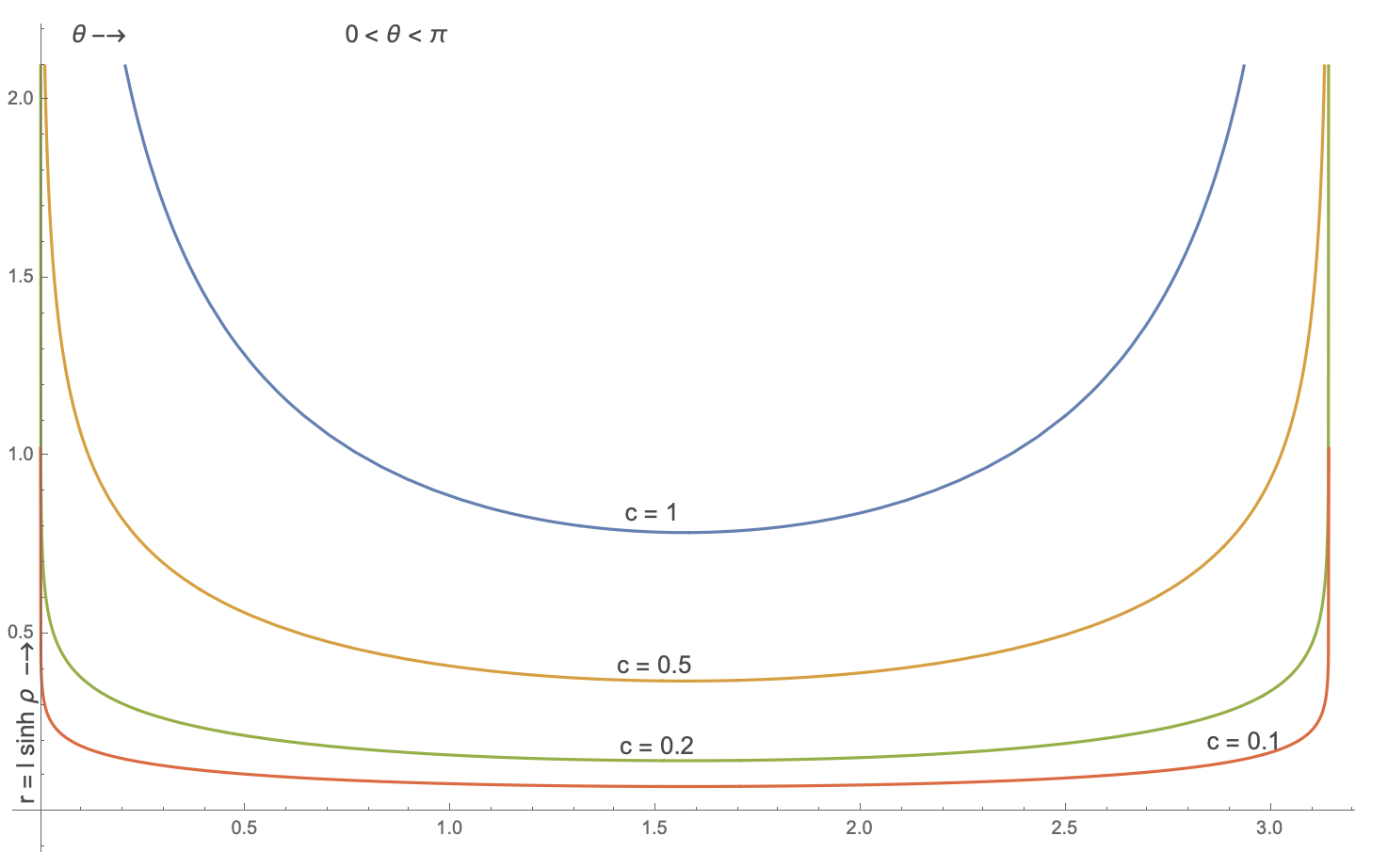}\end{center}
		\caption{{ \label{profileD5}In this figure, we plot the profile of the probe D5 in the $r$-$\theta$ directions($r = l \sinh \rho$) at a fixed $\alpha$ and for different values of the constant parameter $c$.
				\footnotesize   
		}}
	\end{figure}
\end{center}

\subsubsection{Case Example: $\Phi_0 \Phi_1 Z_3^2 = c_1\,;$ $Z_2 = 0 \,;$} \label{hangingprobe}

In terms of the real coordinates, these conditions take the form
\begin{align}\label{simpleD5}
\sinh 2 \rho \, \sin \theta \, \sin^2 \alpha\, =\, c \,, \qquad  \xi_3 \, =\, - \frac12 \left( \phi_0 \, + \, \phi_1\right) \,, \qquad \beta \, = \, \frac{\pi}2
\end{align}
We work with a static gauge condition where we consider the following identification for the worldvolume coordinates
\begin{align}
    \sigma_0 \sim t \,\quad \sigma_1 \sim \theta \,\quad \sigma_2 \sim \phi_1 \,\quad \sigma_3 \sim \phi_2 \,\quad \sigma_4 \sim \alpha \,\quad \sigma_5 \sim \xi_1
\end{align}
The induced metric on the resulting D5 worldvolume is
\begin{align}
    ds^2 \bigg|_{D5} =&  - \cosh^2 \rho \, d\phi_0^2 \, +  \frac{c^2}{\sin^4 \alpha \, \sin^2 \theta + \frac{c^2}4 } \left( \frac{\cot \theta}2 \, d\theta \, + \, \cot \alpha \, d\alpha \right)^2 \,\cr
     +&  \frac{c^2}{4 \sin^2 \theta \sin^4 \alpha} \left( d\theta^2 \, + \, \cos^2 \theta \, d\phi_1^2 \, + \sin^2 \theta \, d\phi_2^2\right) + d\alpha^2  + \, \cos^2 \alpha   d\xi_1^2 + \, \frac{\sin^2 \alpha}4 \left(d\phi_0 + d\phi_1 \right)^2 \cr
\end{align}
By taking the large $AdS$ radius limit($r$ $\rightarrow \, \infty$), we can infer that the dual defect would have a topology of $\mathbb{R} \times \tilde{S}^2$
\begin{align}
    ds^2 \bigg|_{D5;\, \rho \rightarrow \infty } \, \sim & \, \frac{c^2}{\sin^2 \theta \sin^4 \alpha} \left( d\phi_0^2 \, + \,  d\theta^2 \, + \, \cos^2 \theta \, d\phi_1^2 \, \right)
\end{align}
We can also plot the profile of this brane in the $r$-$\theta$ directions at a fixed value of $\alpha$ in figure \ref{profileD5}. In this plot we find that the profile the probe D5 adjusts towards the boundary direction as we take the larger and larger value of the constant real parameter $c$ in the embedding condition in equation \eqref{simpleD5}.

\subsubsection*{Action of the probe D5}

The action of this case example will be given by the following
\begin{align}
    S\bigg|_{D5} \, = \,T_5 \int d^6 \sigma \sqrt{- \text{det} \, g} \, + \, T_5 \int d^6 \sigma \, L_{WZ}
 \end{align}
In the limit where we consider the parameter $c$ to be a large value such that the probe D5 sticks out slightly away from the boundary in the $r$ direction the probe D5 for this choice of $c$ is heavy and therefore the dynamical part contained in the $L_{DBI}$ part of the action does not contribute and the action effectively becomes
\begin{align}
    S\bigg|_{D5} \, \approx \, T_5 \int d^6 \sigma \, L_{WZ}
 \end{align}
This action is essentially topological, with only the Wess-Zumino term present, given by
\begin{align}
L_{WZ} \, = \, \text{d} a \, \wedge \, P[C^{(4)}] 
\end{align}
where $P[C^4]$ is the pullback of the four-form potential coming from the self-dual field strength solution given in \eqref{selfdual5form} 
\begin{align}
    F^{(5)} \, =\, \left(1 \, + \, \star_{10} \right) \text{d}  C^{(4)}
\end{align}
Since this Chern-Simons term in the D5 action comes from the anomaly polynomial on an auxiliary seven-manifold $\mathcal{N}_7$ for which this D5 is a boundary($D_5$ = $\partial \mathcal{N}_7$), for the closed form $F^{(5)}$ this action is equivalent to the following worldvolume integral
\begin{align}
    S\bigg|_{D5} \, \approx \, T_5 \int d^6 \sigma \, a \, \wedge \, F^{(5)} 
 \end{align}
Next we consider the 1-form worldvolume gauge field $a$ to be of the following form
\begin{align}
a \, \propto \,  \left(  d\phi \, + \, d\psi \right) 
\end{align}
which will contribute with a non-trivial holonomy factor multiplication in the action
\begin{align}
    S\bigg|_{D5} \, \approx & \, T_5 \int_{\gamma}a \, \int_{\Sigma_3 \times \Sigma_2} \, F^{(5)} \cr
    = & \, T_5 \, \alpha_{\gamma} \, \int_{\Sigma_3} \, \star_{5} \, F^{(2)} \int_{\Sigma_2} d\text{vol}_{\Sigma_2}
 \end{align}
where $\Sigma_3$ $\subset$ deformed $AdS_5$ and $\Sigma_2$ $\subset$ compact $X^5$(which is equivalent to $S^2$ sphere). And the action evaluates to
\begin{align}
    S\bigg|_{D5} \, \approx  \, T_5 \, \alpha_{\gamma} \, \int_{\Sigma_3} \, \star_{5} \, F^{(2)} \int_{S^2} d\text{vol}_{\Sigma_2}
 \end{align}
where $\alpha_{\gamma}$ is the holonomy parameter. The action contains a quadratically divergent piece in $\frac1{\sin \theta}$, which can be handled by adding a suitable counterterm proportional to $r(\theta)$ at the radial boundary, plus a finite term giving us the following for the effective action value
\begin{align}
    S_{\text{eff.}}\bigg|_{D5} \, =  \, T_5 \, \alpha_{\gamma} \, 96 \sqrt{3}\, \pi^5 \, f \, l^6 
 \end{align}
\subsection{Probe D5 anti-brane solution}\label{hangingantiD5}

The following D5 solution is the anti-brane example in this type 2b background solution
\begin{equation}\label{antiD5}
    \overline{F}^I \left(\bar{\Phi}_0, \bar{\Phi}_1, \bar{\Phi}_2, \bar{Z}_1, \bar{Z}_2, \bar{Z}_3 \right) \, = \, 0 \, \qquad \qquad \left(\text{with} \,\, I= 1,2\right),
\end{equation}
which are two arbitrary holomorphic functional conditions satisfying the scaling condition
\begin{equation}
    \sum_{i=0}^2 \partial_{\phi_i} \bar{F}^I - \sum_{i=1}^3 \partial_{\xi_i} \bar{F}^I \, = \, 0 \,.
\end{equation}
This solution is derived by solving the $\kappa$-symmetry equation with a minus sign
\begin{equation}\label{kappasymmetryeqn}
    \Gamma_{\kappa} \, \epsilon \, = \, - \sqrt{- \text{det} \, h} \, \epsilon
\end{equation}
To get the above general anti-$\overline{\text{D5}}$ brane solution, we considered the reversal in the orientation of the $\widetilde{AdS_5}$ coordinates
\begin{align}
t \rightarrow - t \quad r \rightarrow -r  \quad \theta \rightarrow - \theta \quad \phi_1 \rightarrow -\phi_1 \quad \phi_2 \rightarrow -\phi_2
\end{align}
or equivalently the pullbacks pick up a minus sign
\begin{align}
\mathfrak{e}^a_i \, = \, - \, e^a_{\mu} \partial_i X^{\mu}  \qquad (\text{where}\,\, a \in \{0,1,2,3,4\})
\end{align}
The 6-form volume form on the anti $\overline{\text{D5}}$ is 
\begin{equation}
    \text{dvol}_6 \, = \, \mathfrak{e}^{09} \wedge \left( \omega \, + \, \widetilde{\omega}  \right) \wedge \left( \omega \, + \, \widetilde{\omega}  \right)
\end{equation}
with the definitions of the 2-forms $\omega$ and $\widetilde{\omega}$ are the same as given in equation \eqref{kaehlerforms}.
%
\subsubsection*{Case example}
In the next section, we will consider the simpler anti $\overline{\text{D5}}$ brane with the embedding condition
\begin{align}\label{antiD5example}
\overline{\Phi}_0 \overline{\Phi}_1 \left( \overline{Z}_3 \right)^{-2} = \overline{c}_1 \qquad \overline{Z}_2 \, =\, 0
\end{align}
whose worldvolume overlaps with the worldvolume of the D5 brane example of the previous subsection \ref{hangingprobe} with opposite orientation.

\section{Probe D3 solutions with compact worldvolume}

In this section we want to discuss some probe D3 solutions that are supersymmetric in the background of previous section \ref{deformedAdS}. The solution is very general 
\begin{equation}
    F^I \left(\Phi_0, \Phi_1, \Phi_2, Z_1, Z_2, Z_3 \right) \, = \, 0 \, \qquad \qquad \left(\text{with} \,\, I= 1,2,3\right),
\end{equation}
which are three arbitrary holomorphic functional conditions, also satisfying the scaling condition given below
\begin{equation}
    \sum_{i=0}^2 \partial_{\phi_i} F^I + \sum_{i=1}^3 \partial_{\xi_i} F^I \, = \, 0 \,.
\end{equation}
This general solution has already been derived previously in \cite{AS:2008} in the $AdS_5 \times S^5$ and has been analyzed in good detail in the context of codimension-2 defects in \cite{AGS:2020}. We will be focusing on the subset of the BPS condition which gives the embedding condition of giant graviton D3 branes of \cite{Mikhailov:2000, Grisaru:2000zn, McGreevy:2000cw}. The holographic duals of the giants in the $\mathcal{N}=4$ SYM are the heavy determinant operators which may become charged under the $U(1)$ 0-form global symmetry of interest. 

The maximal supersymmetric solution here wraps an $S^3$ sphere inside the $S^5$ and a time like curve in the $AdS$. In this paper we are interested in the solution of the form
\begin{align}\label{giantD3branes}
Z_1 \Phi_0 \, = \, c_2 \, e^{i \xi^0}  \qquad \frac{\Phi_1}{\Phi_0} \, = \, c_3 \qquad \frac{\Phi_2}{\Phi_0} \, = \, c_4
\end{align}
where $c_2$, $\xi^0$ are real constant parameters and $c_3$, $c_4$ are complex constants.
In holography, these D3 giant branes are dual to the heavy determinant operators of the form
\begin{align}
\text{det} \, Z \, = \, \frac1{N!} \varepsilon^{i_ii_2 \ldots i_N} \varepsilon_{j_i j_2 \ldots j_N} Z_{i_ii_2 \ldots i_N}^{j_i j_2 \ldots j_N}
\end{align}
where $Z$ is complex scalar in the adjoint representation of $SU(N)$. \\

\noindent
Our proposal in this paper is that these heavy operators will carry a nontrivial charge under the symmetry operators dual to the flattened D5 branes we discussed in the previous section.

\subsubsection*{Action of the probe D3}

In terms of real coordinates, the conditions in \eqref{giantD3branes} look as the following

\begin{align}
 r \, = \, l \sinh \rho = \frac{|c_3|}{\sqrt{|c_3|^2 -1}} &\qquad \alpha \, = \, 0 \qquad \xi_1 \, + \, \phi_0 \, =\, \xi^0 \cr
 \phi_1 \, - \, \phi_0 \, = \, Arg[c_3] & \qquad \phi_2 \, - \, \phi_0 \, = \, Arg[c_4] \qquad 
 \theta \, = \, \text{constant}
\end{align}
The induced metric on the worldvolume of this giant graviton brane is the following
\begin{align}
    ds^2 \bigg|_{D3} \, =\, - dt^2 \, + \, l^2 \left( d\beta^2  \, + \, \sin^2 \beta \, d\xi_2^2 \, + \, \cos^2 \beta \, d\xi^2   \right)
\end{align}
This brane is located at a fixed radial distance away from the boundary and therefore by tuning the parameter $|c_3|$ we can shift it closer and closer to the boundary. And for the purpose of discussion in this article, we would like to keep it between the bulge of the \textit{hanging} D5 solution of the figure \ref{profileD5} in the previous section(with the constant $c$ large) and the boundary. And therefore, the limit where the parameter $|c_3|$ will be relevant for us.

\noindent
The action of this giant graviton D3 brane will be 
\begin{align}
    S\bigg|_{D3} \, =& \,  \int d^4\sigma \,  L_{DBI} \, + \,  \int d^4\sigma \, L_{WZ} \cr
    =& \, T_{D3} \int d^4\sigma \,  \sqrt{- \text{det}[g \, + \, \mathbf{F}]} \, + \, T_{D3} \int d^4\sigma \, P[C^{(4)}]
\end{align}
where $g$ is the induced metric on the D3 brane and $\mathbf{F}$ is the field strength due to the $U(1)$ worldvolume gauge field, which can be activated on the brane. \\
%
%

\noindent
The $U(1)$ field $a$ that we consider on this D3 will be of this form 
\begin{align}
a \, = \, d\phi_0 - d\xi_1
\end{align}
for which the field strength $\mathbf{F}$ vanishes everywhere.
\begin{align}
    S\bigg|_{D3} \, =& \, T_{D3} \int d^4\sigma \,  \sqrt{- \, \text{det}\, g } \, + \, \int d^4\sigma \, P[C^{(4)}]
\end{align}
\subsection{Charge measurement via holography}\label{chargemeasurement}
\begin{figure}
     \centering
     \begin{subfigure}[b]{0.495\textwidth}
         \centering
         \includegraphics[width=\textwidth]{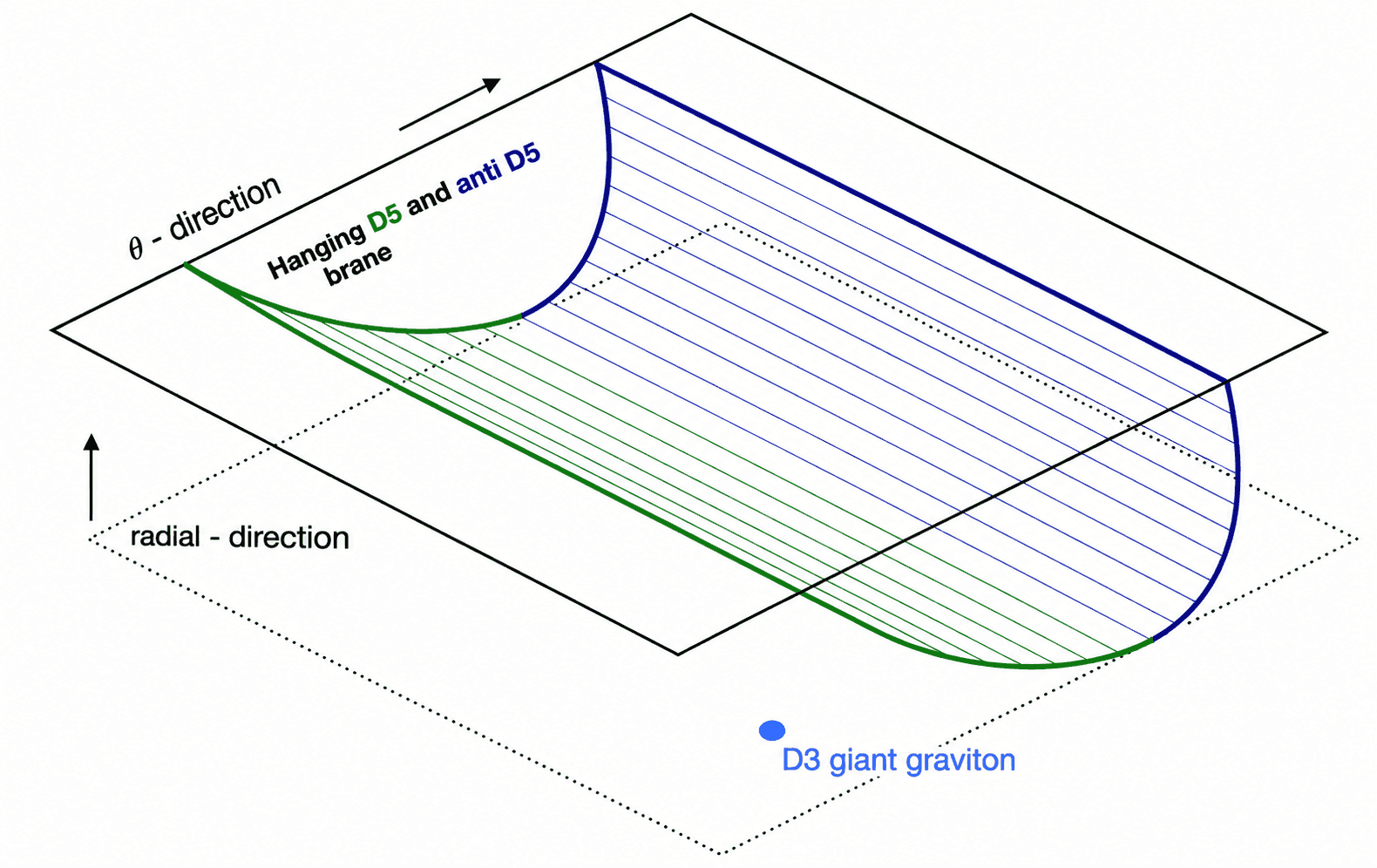}
         \label{charge measurement 1}
     \end{subfigure}
     \hfill
     \begin{subfigure}[b]{0.495\textwidth}
         \centering
         \includegraphics[width=\textwidth]{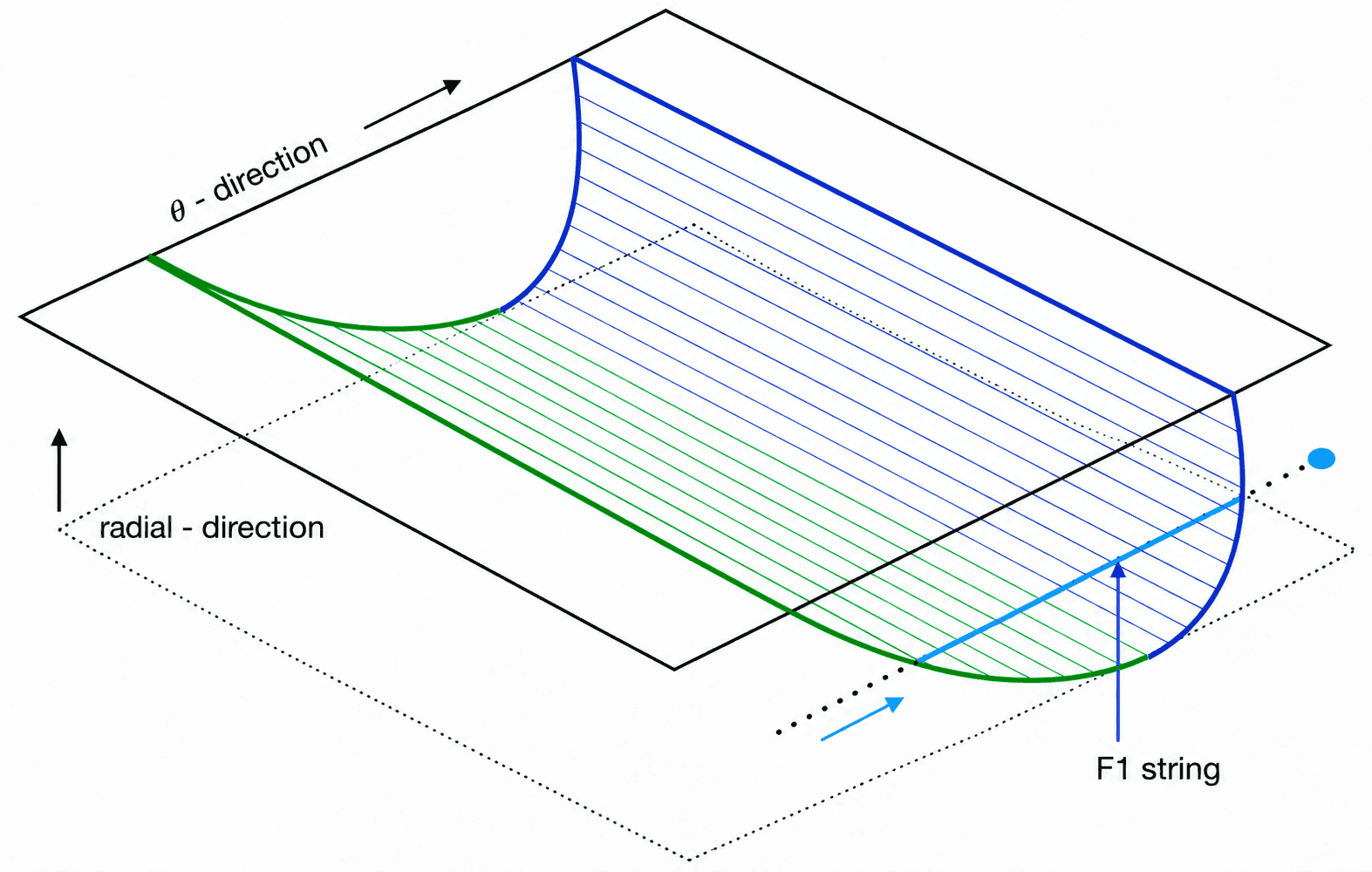}
         \label{charge measurement 2}
     \end{subfigure}
        \caption{{\small Charge measurement: the figures depict the positioning of the probe D5 and anti-$\overline{\text{D5}}$ brane, placed alongside, overlapping, and hanging from the radial boundary. Their combined profile is stretched along $\theta$ direction and the other three spacetime directions. A giant graviton D3 brane(represented by a blue dot) is placed between the boundary and the maximum bulge of the D5-$\overline{\text{D5}}$(in the radial direction). In the right figure, we show the relative movement between the giant graviton brane and the hanging D5-$\overline{\text{D5}}$ pair in the $\theta$ direction, which creates a fundamental string due to the Hanany-Witten effect.}}
        \label{charge measurement}
\end{figure}
Following the references \cite{Bahetal:2025, Bahetal:2026}, we show how holographically, the charge measurement of the heavy determinant operator in the boundary SYM theory can be done by changing the relative position of the hanging D5 brane solution and the giant graviton D3 brane. The hanging probe D5 brane solution of subsection \ref{hangingprobe} is considered to be placed alongside, overlapping with its anti $\overline{\text{D5}}$ brane counterpart solution of subsection \ref{hangingantiD5}, and together the combination is dual to topological defects of a $U(1)$ 0-form symmetry. In the $r-\theta$ plane, the profile of the hanging D5-$\overline{\text{D5}}$ brane pair trace out a curve whose ends are anchored at the radial boundary and it bulges towards the radial center as seen in figure \ref{charge measurement}. But the duality of D5-$\overline{\text{D5}}$ branes with the $U(1)$ symmetry defect is only true when this bulge stays close to the radial boundary i.e. when we consider the real parameter $c$ in equation \eqref{simpleD5} of subsection \ref{hangingprobe} to be large.

Since the giant solution traces a time-like trajectory in the bulk we consider it to be placed somewhere between the boundary and the maximum radial position of the bulging point of the hanging D5 brane as shown in the figure \ref{charge measurement}. The giant gravition wraps an $S^3$ sphere in the compact directions $X_5$ which is completely orthogonal to the two directions($\Sigma_2$) the hanging branes wrap in $X_5$. The brane table below depicts the orthogonal and parallel directions of the D5-$\overline{\text{D5}}$ pair and the D3 giant.

 \begin{table}[H]
	\begin{center}
		\begin{tabular}{|c|c|c|c|c|c|c|c|c|c|c|}
			\hline
			 $\qquad \qquad \,\,\,\,\,$&  $\phi_0$ &  $\phi_1$& $\phi_2$& $\,r\,$& $\theta$& $\alpha$& $\beta$  & $\xi_1$ & $\xi_2$ & $\xi_3$ \cr
			\hline
        \end{tabular}   
        \begin{tabular}{|c|c|c|c|c c|c|c|c|c|c|}
			\hline
        D5-$\overline{\text{D5}}$ pair &  $\,\times\,$&   $\times$& $\,\times$& $\,\,\,\times$ &  & $\times$ & $-$ & $\times$ &$-$ & $-$  \cr
            \hline 
        \end{tabular}  
        \begin{tabular}{|c|c|c|c|c|c|c|c|c|c|c|}    
            \hline
$\,\,\,$D3 giant$\,\,\,\,$ &  $\,\times\,$& $-$& $-$& $-$& $-$ & $-$ &  $\times$ & $-$ & $\times$ & $\times$ \cr
			\hline 
		\end{tabular}
        \begin{tabular}{|c|c c c|c|c|c|c|c|c|c|}
		\hline
			$\,$F1 string\,$\,\,\,\,$& $\,\,\,\,\,\,\,\,\,\,\,\times$& &   & $-$ &  $\times$ &$-$ & $-$ &  $-$& $-$& $-$\cr
                \hline
		\end{tabular}
	\end{center}
	\caption{Table showing the orthogonal directions of worldvolume of D5-$\overline{\text{D5}}$ pair, D3 giant and the newly created fundamental string, } 
	\label{M5construction}
\end{table}

To explain the $U(1)$ symmetry charge of the boundary operator, we consider the relative movement between the giant graviton brane and the hanging D5-$\overline{\text{D5}}$ pair in the $\theta$ direction. When the D5 brane is passed across the D3 along the dotted line trajectory(in figure \ref{charge measurement}), a fundamental string is created between the two types of branes due to the Hanany-Witten effect \cite{HananyWitten:96}. After this, the anti-$\overline{\text{D5}}$ is also made to pass through the D3 so that it ends up at the location alongside the D5 brane, overlapping with it again. After this relative shift between the D3 giant and the D5-$\overline{\text{D5}}$ pair, a fundamental string is created along the $\theta$ direction, stretched between the two points on the D5-$\overline{\text{D5}}$ pair where they intersected with the passing D3 brane(as shown in figure \ref{charge measurement} using a dark blue line). The two endpoints of the F1 string couple to the $U(1)$ Chan-Paton gauge field $a$ on the D5 brane worldvolume, and it's other single direction  wraps along the $\gamma$ curve on the D5.

This Hannay-Witten transition process would lead to the effective action acquiring a phase factor given by
\begin{align}
    \exp\left[ 2 \pi i \left( \Sigma_2 \cdot S^3 \right) \int_{\gamma} a  \right] \, =\, \exp\left[ 2 \pi i \, \alpha_{\gamma} \left( \Sigma_2 \cdot S^3 \right)  \right]
\end{align}
where $\left( \Sigma_2 \cdot S^3 \right)$ is the intersection number between the submanifolds $\Sigma_2$ and $S^3$ wrapped by the hanging D5-anti-$\overline{\text{D5}}$ pair and the giant D3, respectively, in the compact 5-directions on $X_5$ of the 10d spacetime. Our proposal in this work is that this phase factor corresponds to the charge acquired by the holographically dual determinant operator in the boundary $\mathcal{N}=4$ SYM theory under the $U(1)$ global 0-form symmetry. 

\section{Conclusion}

In this work, we have derived some probe D5 brane solution by solving the $\kappa$-symmetry equation of a type 2b supergravity background which is a deformation of $AdS_5 \times S^5$ with the $U(1)$ 1-form gauge field turned on. Our solution for D5 embedding conditions in section \ref{generalprobeD5} was very general and were expressed in terms of two arbitrary holomorphic functions in \eqref{generalprobeD5solution}. From these general conditions, we focus on a simpler case of a D5 worldvolume which is noncompact and extends along the radial direction and whose profile can be made to take the U-shaped hanging profile anchored from the radial boundary. This was achieved by taking the real parameter $c$ in the embedding condition \eqref{simpleD5} to be large. When we placed this probe D5 in \eqref{simpleD5} alongside its anti-$\overline{\text{D5}}$-brane counterpart from \eqref{antiD5example}, overlaping in 4 spacelike directions, the combination became suitable to discuss the holography of the codimension-1 topological defect associated with a $U(1)$ global symmetry in the boundary field theory. In the last section, we show how one can explain the charge measurement of the heavy determinant operators under the global symmetry of the boundary theory in the holographic setup, by doing a relative movement between D5-$\overline{\text{D5}}$ and the giant graviton in the bulk along the $\theta$ direction. This relative shift between the two types of branes induces a Hanany-Witten transtion effect giving us the value of the charge measured under this procedure.

In the future, we would like to study the probe solution in section \ref{section2} in a bit more detail without assuming any large-value limit for the embedding parameter $c$. It would be interesting to compute the correlation functions in the presence of the dual codimension-1 defect and to analyse how tuning $c$ affects the results.\\

\textbf{Acknowledgement:} The Author would like to thank the organisers of the program \textit{Generalised symmetries and anomalies in quantum phases of matter 2026} at ICTS-TIFR and the organisers of the event \textit{Global Categorical Symmetries 2026} held at IHP, where ideas in this work were partially discussed. The Author would also like to thank the Institute of Mathematical Sciences, Chennai, for the hospitality last year.

\providecommand{\href}[2]{#2}\begingroup\raggedright\endgroup

\end{document}